# Microwave-assisted cross-polarization of nuclear spin ensembles from optically-pumped nitrogen-vacancy centers in diamond

*F. Shagieva, S. Zaiser, P. Neumann, D.B.R. Dasari, R. Stöhr, A. Denisenko, R. Reuter, C.A. Meriles, and J. Wrachtrup\**

[1]*Universität Stuttgart, 3. Physikalisches Institut, Stuttgart, Germany.*
[2]*CUNY – City College of New York, New York, 10031 NY, USA.*

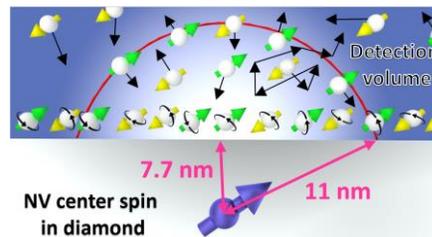

**ABSTRACT:** The ability to optically initialize the electronic spin of the nitrogen-vacancy (NV) center in diamond has long been considered a valuable resource to enhance the polarization of neighboring nuclei, but efficient polarization transfer to spin species outside the diamond crystal has proven challenging. Here we demonstrate variable-magnetic-field, microwave-enabled cross-polarization from the NV electronic spin to protons in a model viscous fluid in contact with the diamond surface. Slight changes in the cross-relaxation rate as a function of the wait time between successive repetitions of the transfer protocol suggest slower molecular diffusion near the diamond surface compared to that in bulk, an observation consistent with present models of the microscopic structure of a fluid close to a solid interface.

The practice of nuclear magnetic resonance (NMR) presently encompasses multiple disciplines spanning the analytical and medical sciences, biochemistry, environmental monitoring, and well logging, to mention just a few. A limitation common to all these applications, however, is the sample spin polarization, which is typically a minute fraction of the possible maximum ($10^{-5}$ at 10 T under ambient conditions). Sensitivity restrictions set a limit on the minimum amount of sample that can be detected (a concern when analyzing mass-limited samples or rare molecular moieties in solution), and result in longer acquisition times. In particular, because the signal-to-noise ratio of an NMR measurement grows with the square root of the number of repeats, just a two-fold enhancement of the sample polarization leads to a four-fold reduction of the acquisition time. Resorting to cryogenic temperatures or stronger magnets are the most obvious routes to enhanced spin polarization (and hence improved detection sensitivity), but sample freezing is often impractical (consider, e.g., living organisms) and large magnets tend to be disproportionately expensive.

Adding to the existing library of free-radical-based dynamic nuclear polarization (DNP) schemes[1], the use of optically polarized paramagnetic defects in wide bandgap semiconductors is emerging as an alternative polarization enhancement route of growing interest. An example of prominent importance is the negatively-charged nitrogen-vacancy (NV[-]) center in diamond, a spin-1 system formed by a substitutional nitrogen and an adjacent vacancy.

Green illumination efficiently pumps NV[-] into the $m_S = 0$ state of the ground state spin-triplet via spin-selective intersystem crossing[2]. This feature has already been exploited to demonstrate record levels of room temperature [13]C spin polarization at low[3-8] (i.e., <100 mT) and intermediate[9] (i.e., <300 mT) magnetic fields; optically pumped [13]C polarization has also been observed at 7 T under cryogenic conditions[10]. Recent studies based on other defects in diamond[11] or wide-bandgap host crystals other than diamond[12] have led to comparable results.

In spite of this progress, dynamic polarization of electron or nuclear spin species (within or outside the crystal lattice) through optically pumped NVs is still the subject of much research. Contributing to this trend are the many alternative polarization transfer mechanisms (activated resonantly at select magnetic fields[3-7] or via the use of microwave (mw)[8,9]), and the complexity of the dynamics at play (particularly if the NV[-] axis and the direction of the external magnetic field do not coincide[13]). For example, cross-relaxation of optically-pumped shallow NVs interacting with the protons of an organic substance on the diamond surface has been recently observed near the NV[-] ground state level anti-crossing (~100 mT), where one of the allowed NV[-] spin transitions approximately matches the proton Zeeman splitting[14].

Here we extend this strategy to magnetic fields other than 100 mT via a mw-mediated polarization transfer scheme where the amplitude of the mw field driving the NV is chosen so that the Rabi and Zeeman frequencies of the electronic and nuclear spins coincide[15]. Cross polarization to ¹³C spins within the diamond crystal has already been demonstrated via this scheme[3,16], but extensions to outside spins have proven challenging, partly because of the sensitivity to mw amplitude fluctuations in the limit of weak hyperfine couplings. The present work overcomes this problem to demonstrate cross-relaxation of individual NV⁻ spins to protons in a model fluid at variable magnetic fields[17]. By exploiting the dependence of the polarization transfer efficiency on the local dynamics of the target system, we show that molecular diffusion is substantially slower than in the bulk fluid, thus supporting current theoretical views on the structure of a liquid near a solid interface.

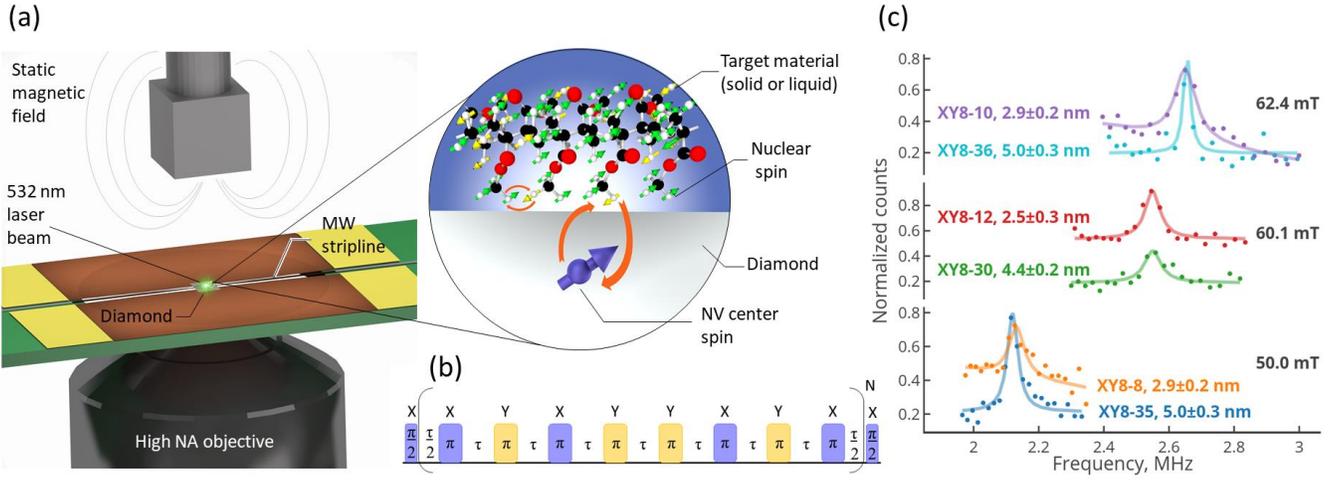

**Figure 1 | NV⁻-assisted detection of proton spins on the diamond surface.** (a) Working geometry. We engineer shallow NVs via a fabrication process involving diamond overgrowth, nitrogen implantation, and sample annealing (see main text). (b) To determine the NV depth, we implement an XY8 pulse train and record the NV⁻ spin response as a function of the inter-pulse separation τ measuring the NMR spectrum of a sample on top of the diamond. (c) ¹H NMR spectra as obtained upon application of the protocol in (b) for different magnetic fields. In the above experiments, all mw pulses act resonantly on the NV⁻ $m_S = 0 \leftrightarrow m_S = -1$ transition. NA: Numerical aperture (1.35).

Fig. 1a shows a schematic of our working geometry. In our experiments, we use a commercial electronic grade [100] crystal from E6 overgrown in-house with a sacrificial layer of a boron-enriched diamond via mw-assisted chemical vapour deposition[18]. We create NV centers via ¹⁵N ion implantation followed by sample annealing; removal of the boron-rich layer is subsequently carried out through chemical etching (see Section 1 in the Supplementary Material). Throughout our experiments, we probe the NV⁻ spin via optically detected magnetic resonance (ODMR). To determine the NV depth, we implement an XY8 multi-pulse train (Fig. 1b) and monitor the NV⁻ response as we change the inter-pulse separation; a signal reduction occurs when the latter matches half the proton spin Larmor period[17]. By assuming a proton density of 50 nm⁻³, the NV⁻ depth — ranging from 2.5±0.3 nm to 5.0±0.3 nm in the examples of Fig. 1b — can be extracted from the root mean square value of the acting magnetic field, in turn, connected to the measured ¹H NMR peak amplitude[18]. On average, we find these NVs show relatively long spin-lattice relaxation ($T_1$) and coherence ($T_2$) lifetimes, reaching up to 7.7 ms and 67.2 μs, respectively.

Fig. 2a shows the cross-polarization scheme we use: Upon spin initialization via green laser illumination (532 nm, 3 μs), we align the NV⁻ spin along the mw-field via a π/2-pulse followed by a 'locking' pulse in quadrature; the resulting NV⁻ spin polarization is determined via a projection π/2-pulse followed by a 3 μs readout pulse of green light. When the mw amplitude is adjusted so that the electronic Rabi splitting matches the nuclear spin Zeeman splitting, off-diagonal terms $A_\perp$ in the hyperfine coupling active during the spin locking pulse drive the coherent transfer of spin polarization from NV⁻ into proximal nuclei at a rate $\sqrt{A_\perp^2 + (\Omega - \omega_L)^2}$, and the magnitude of the polarization transferred (determined by the depth of decay curves in Fig. 2c – 2e) is $A_\perp^2/(A_\perp^2 + (\Omega - \omega_L)^2)$. We will show later that this transfer rate is further affected by the relaxation rate of the NV spin and the diffusion in external spins. Clearly, the transfer is maximized at the resonant condition i.e., more formally, we write

$$\Omega = \gamma_{\text{NV}} B_1 = \omega_L = \gamma_{\text{H}} B_0 , \qquad (1)$$

where Ω is the NV⁻ spin Rabi frequency, $\gamma_{\text{NV}}$ ($\gamma_{\text{H}}$) is the NV⁻ (¹H) gyromagnetic ratio, and $B_1$ ($B_0$) is the mw (Zeeman) field. An adaptation of the Hartmann-Hahn condition[19,20], this form of 'rotating frame' cross-polarization has already found extensive use in traditional (i.e., macroscopic) inductively-detected magnetic resonance, where it is best known as NOVEL[15,21] (Nuclear spin Orientation Via Electron spin Locking).

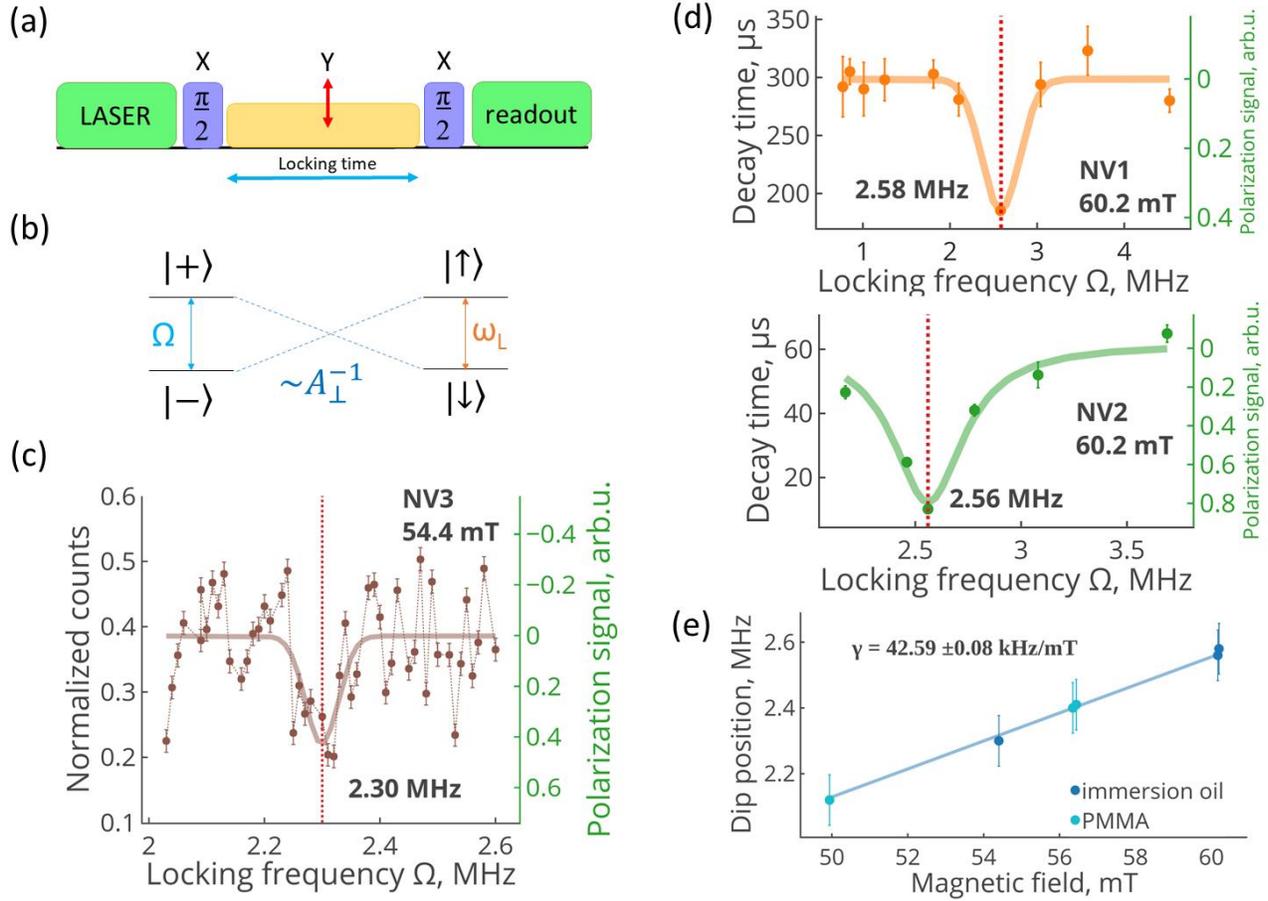

**Figure 2 | Resonant NV⁻ spin cross relaxation to ¹H spins in immersion oil.** (a) Schematics of the NOVEL protocol. We measure the NV⁻ spin polarization after a "spin-locking" pulse of variable duration and/or amplitude. (b) NV⁻ spin cross relaxation to the nuclear spin bath takes place when the Rabi frequency matches the nuclear spin Zeeman splitting; the transfer time is dictated by the inverse of the hyperfine coupling constant $A_\perp$. (c) NV⁻ spin signal (depth $7.7 \pm 0.4$ nm) after a NOVEL sequence of fixed spin-locking duration (150 µs) and variable mw amplitude (expressed as the NV⁻ spin Rabi frequency). (d) NV⁻ spin relaxation time $T_{1\rho}$ during spin-locking as a function of the Rabi field for two shallow NVs ($3.9 \pm 0.2$ nm and $2.6 \pm 0.1$ nm deep, top and bottom, respectively). The solid line in (c) and (d) is a Lorentzian fit whereas the vertical dashed line indicates the ¹H spin resonance frequency at the applied magnetic field (60.2 mT). (e) Rabi frequency corresponding to the $T_{1\rho}$ minimum as a function of the magnetic field; from the linear fit (solid line), we extract a rate $\gamma = 42.59 \pm 0.08$ kHz/mT, consistent with the ¹H gyromagnetic ratio.

In our experiments, we first record the NOVEL signal from individual shallow NVs brought in contact with proton spins in an organic fluid on the diamond surface. As a proof of principle, we use the immersion oil (Fluka Analytical 10976) used to optimally couple the NV⁻ fluorescence to the optical objective; we will show later that the relatively high viscosity of this fluid — and, correspondingly, its slow molecular diffusion — plays a key role in enabling an efficient polarization transfer. Fig. 2c shows the measured NV⁻ polarization after a fixed spin-locking time (150 µs) as a function of the mw field amplitude: We find greater signals for Rabi frequencies both above and below the proton Zeeman frequency (dashed vertical line), thus confirming the resonant nature of the transfer. A similar dependence can be seen for the NV⁻ spin-locking relaxation time $T_{1\rho}$, presented for two example NVs in Fig. 2d. As we change the applied magnetic field, the dip position shifts at a rate $\gamma = 42.59 \pm 0.08$ kHz/mT (Fig. 2e), coincident with the ¹H gyromagnetic ratio ($\gamma_H = 42.576$ kHz/mT).

We now turn to the dynamics of the polarization process. A direct comparison of the temporal evolution of the NV⁻ signal with or without Hartmann-Hahn matching is presented in Fig. 3a (orange and green traces, respectively). In both cases, we observe an exponential decay of the NV⁻ spin polarization as a function of locking time, though the on-resonance time constant is considerably shorter, a signature of NV⁻ spin cross-relaxation with the proton bath. Calculating the difference between both curves (blue trace in Fig. 3a), we use the slope in the linear response at early times (dotted red line in Fig. 3a) to approximate the cross-polarization time $T'_{CP} = 120$ μs, where the prime highlights the difference with the 'true' cross-polarization time $T_{CP}$ observed in the limit of an infinite wait time $T_W$ between successive repeats (see below). Assuming for now the difference between the two is not large, we obtain a crude estimate of the distance to the detection volume via the approximate formula[22]

$$\frac{2\pi}{T'_{CP}} \cong \frac{\mu_0}{4\pi} \frac{\gamma_{NV}\gamma_H \hbar}{r_{eff}^3} \sqrt{N} \qquad (2)$$

where $N$ is the number of protons within the detection volume, $\hbar$ is Planck's constant divided by $2\pi$, and $\mu_0$ is the permeability of free space. Using $N \sim 7000$, we get $r_{eff} \sim 10$ nm, consistent with the measured NV⁻ depth (7.7 ± 0.4 nm).

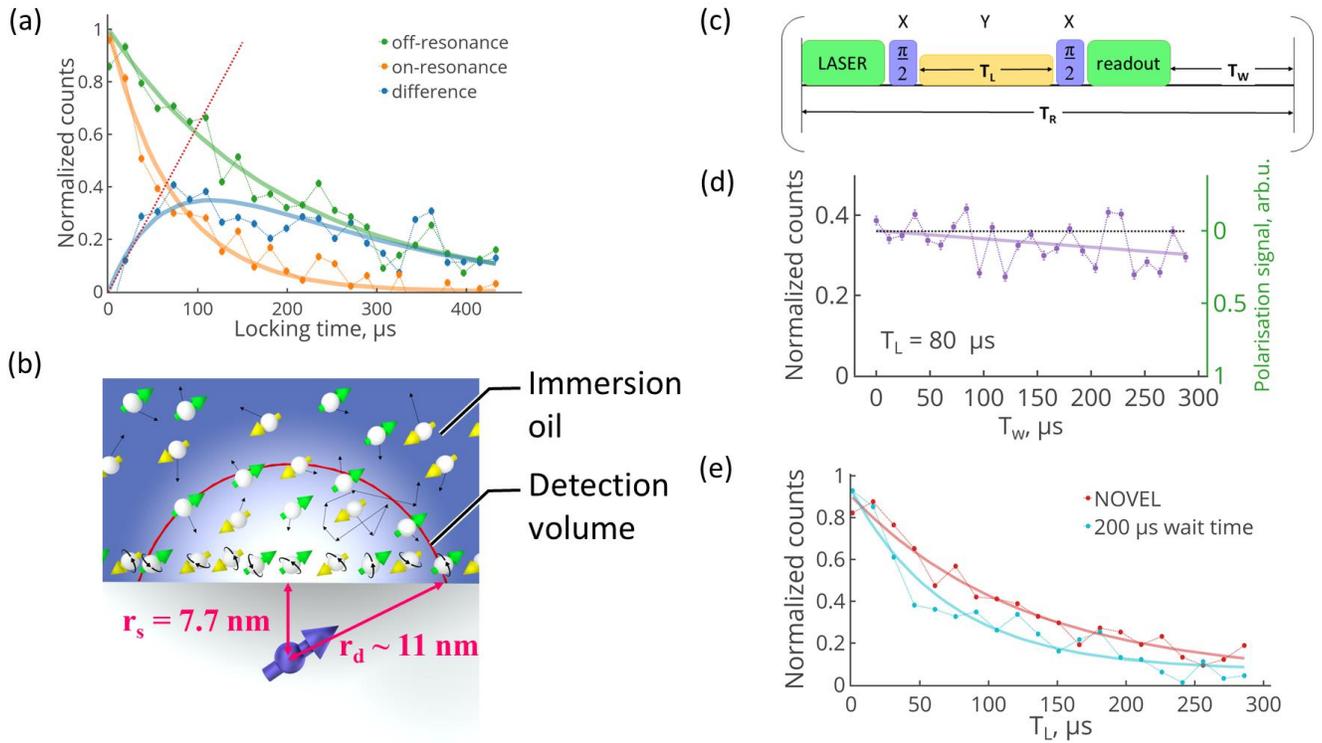

**Figure 3 | Dynamics of NV⁻ cross-relaxation.** (a) NV⁻ spin signal after a NOVEL protocol of variable spin-locking time for Rabi frequencies on- and off-resonance relative to the Larmor frequency of protons in immersion oil (orange and green traces, respectively). To determine the cross-polarization time, we take the difference between the two (blue trace); from the inverse of the slope at early times (dashed line), we find $T'_{CP} = 120$ μs. (b) An NV at a distance $r_S = 7.7$ nm from the surface effectively interacts with all protons within a distance $r_S \sim 11$ nm (detection volume). This geometry leads to a broad distribution of hyperfine couplings preventing the formation of polarization transfer revivals. (c) NOVEL protocol featuring a wait time $T_W$ between successive repeats. (d) NOVEL NV signal as a function of $T_W$ for a fixed spin-locking time $T_L = 80$ μs. (e) Same as in (d) but as a function of $T_L$ for a wait time $T_W \sim 200\ \mu s$ (cyan) and zero wait time $T_W \sim 0$ (red) traces, respectively. In all cases the magnetic field is 54.4 mT; all other conditions as in Fig. 2.

In stark contrast with prior ¹³C NOVEL experiments in diamond[3] (where the NV⁻ evolution during the spin-lock time shows successive maxima and minima), the exponential decay observed herein indicates a monotonic, one-directional polarization transfer process. The latter, of course, stems from the broad distribution of NV⁻–¹H couplings inherent to the present working geometry and, correspondingly, the lack of commensurate frequencies required to produce a 'revival' in the polarization transfer. Molecular diffusion during $T_{CP}$ further contributes to smooth out the NV⁻ signal, though we note that a similar exponential decay was seen for protons in PMMA (see below and Section 3 in the Supplementary Material).

The ability to transfer spin polarization to neighbouring protons from a single NV can be further exploited to gather information on the chemical composition[23], nanoscale structure[24,25], and molecular dynamics[26] near the diamond surface. Of special interest in the case of fluids is how molecules arrange and diffuse near the interface with the solid[27]. Our initial experiments in this direction make use of the protocol in Fig. 3c, where we introduce a variable wait time $T_W$ between successive repetitions of the NOVEL sequence. The rationale is that because spin diffusion and nuclear spin relaxation within the proton bath are relatively slow, a number of repeats comparable to $N$ (i.e., the number of protons in the detection volume) is sufficient to locally saturate the proton bath polarization. Since $N$ is much smaller than the number of repeats needed to attain a reasonable signal-to-noise (SNR) ratio (of order $10^6$ per data point), a polarization transfer blockade takes place unless molecules physically diffuse in and out of this volume on a faster time scale. Indeed, we find for protons in oil that the signal amplitude after a NOVEL sequence of fixed spin-locking time (80 μs) tends to decrease as $T_W$ grows (Fig. 3c); along the same lines, we observe comparatively faster cross-relaxation after the addition of a fixed wait time of 200 μs (Fig. 3e), thus hinting at diffusion processes out of the detection volume with characteristic times longer than $T_{CP}$.

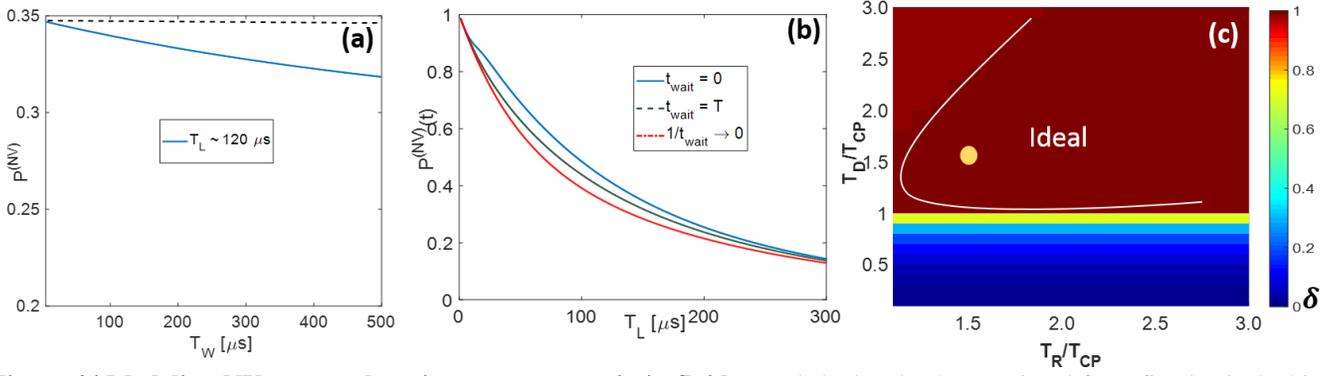

**Figure 4 | Modeling NV⁻ cross-relaxation to protons spin in fluids.** (a) Calculated NOVEL signal for a fixed spin-locking duration as a function of the wait $T_W$ between repeats. (b) Same as in (a) but as a function of the spin-locking time for a short (1 μs) and a long (200 μs) wait time between repeats (blue and orange traces, respectively). For reference, the plot also includes the ideal response in the limit of infinite wait time (red trace). (c) Calculated contrast (see Eq. (4)) as a function of the repetition and diffusion times, $T_R$ and $T_D$, respectively.

To interpret these observations on a more quantitative basis, we model the polarization transfer process through a set of coupled master equations describing the flow of polarization from the NV to protons within the detection volume (which we describe as a single unit block interacting with the NV via an effective coupling of amplitude $T_{CP}^{-1}$). Defining $T_R$ as the time separating two successive repeats, we write

$$\frac{dP^{(NV)}}{dt} = \frac{f(t)}{T_P^{(NV)}}\left(1 - P^{(NV)}\right) - \frac{P^{(NV)}}{T_{SL}^{(NV)}} - \frac{g(t)}{T_{CP}}\left(P^{(NV)} - P^{(H)}\right) \quad (3)$$

and

$$\frac{dP^{(H)}}{dt} = -\frac{P^{(H)}}{T_{SL}^{(H)}} - \frac{P^{(H)}}{T_D^{(H)}} + \frac{g(t)}{NT_{CP}}\left(P^{(NV)} - P^{(H)}\right) \quad (4)$$

Here $P^{(NV)}$ ($P^{(H)}$) denotes the NV⁻ (¹H) spin polarization (within the detection volume), $T_{SL}^{(NV)}$ ($T_{SL}^{(H)}$) is the NV (¹H) spin-lattice relaxation time, $T_D^{(H)}$ is the characteristic time a molecule spends within the detection volume, and $T_P^{(NV)}$ denotes the NV spin pumping time under optical excitation; $f(t)$ ($g(t)$) is a step function (of period $T_R$) equal to 1 when the laser (mw) is on and 0 otherwise.

To numerically solve the above set of equations we assume $T_{SL}^{(H)} \sim 1\,\text{s} \gg T_D^{(H)}$, i.e., the decay of proton polarization in the detection volume is dominated by molecular diffusion. For simplicity, we further assume that the NV⁻ spin-lattice relaxation time $T_{SL}^{(NV)} = 250$ μs is given by the off-resonance NV⁻ spin relaxation time $T_{1\rho}^{(off)}$ (see Fig. 3a). Note that because $T_P^{(NV)} \sim 1$ μs is much shorter than any other time scale, full polarization of the NV⁻ spin is reached almost instantaneously upon laser excitation. With these approximations, we interpret the observations in Figs. 3d and 3e as the result of slow molecular diffusion dynamics near the surface, i.e., $T_D^{(H)} > T_{CP}$. In particular, by assuming $T_{CP} = 100$ μs (see Section 5 in the Supplementary Material), we find the experimentally observed contrast for $T_D^{(H)} \gtrsim 170$ μs. Interestingly, this value exceeds the time molecules spend within the detection volume, $(r_d - r_s)^2/D_M^{(b)} \sim 60$ μs, calculated when assuming that the self-diffusion coefficient of bulk oil, $D_M^{(b)} \sim 0.3$ nm²/μs, remains unchanged near the surface. Our results suggest, therefore, that the dynamics of oil molecules is slower close to the solid-liquid interface, a notion consistent with theoretical models[27,28] and recent experimental observations[26,29].

In applications aimed at the hyperpolarization of target fluids[30,31], an interesting question is the set of conditions — $T_{CP}$, $T_D$, and $T_R$ — required to optimize the polarization transfer. In Fig. 4 we analyze the NV⁻ spin cross-relaxation contrast by calculating the polarization gain by the proton spin when the NV⁻ spin is driven at on- and off-resonance conditions (see Section 5 in the Supplementary Material). For Fig. 4C, we have fixed the cross polarization time $T_{CP}$, and varied the wait time $T_W$, and diffusion time $T_D$. We find that most efficient cross-polarization when $T_D \sim T_R > T_{CP}$ and remains constant for any further increases in $T_R$. On the other hand, at faster diffusion rates ($T_D \ll T_{CP}$), the cross polarization rate averages to zero, thereby reducing the transfer efficiency for any $T_R$. Hence regardless the repetition rate, high diffusive fluids such as water cannot be polarized efficiently through this technique without the formation of a near-surface layer featuring sufficiently slow molecular dynamics.

The discussion above can be formally extended to nuclear spins in solids if $T_D^{(H)}$ is replaced by the spin diffusion time $T_S^{(H)} \sim (r_d - r_s)^2/D_S$, where $D_S \sim (\mu_0/4\pi)(0.1\,\gamma_H^2 \hbar/r_{H-H})$ is the spin diffusion coefficient[32], and $r_{H-H}$ is the inter-nuclear distance. This process, however, is inherently slow: Even for strongly coupled nuclear systems such as protons, we calculate $D_S \sim 2 \times 10^{-4}$ nm²/μs, orders of magnitude smaller than the self-diffusion coefficient of most fluids. The rapid polarization of neighboring nuclei thus prevents further transfer — the so-called 'blockade' — hence leading to negligible NOVEL contrast. Naturally, the contrast can be made non-zero by periodically cycling the sign of the starting NV⁻ spin, though at the expense of a net polarization gain in the nuclear spin reservoir. By the same token, non-zero contrast is expected in the limit of exceptionally short nuclear spin-lattice relaxation times, precisely the condition throughout our NOVEL transfer experiments to protons in a PMMA film (Section 3 in the Supplementary Material).

In summary, we demonstrated microwave-assisted cross polarization from optically-pumped shallow NVs to protons in an organic fluid on the diamond surface. By studying the polarization transfer efficiency as a function of the repetition rate we obtained an estimate for the effective time an oil molecule spends within the detection volume. The corresponding near-surface diffusion coefficient $D_M^{(s)} \lesssim 0.1$ nm²/μs is substantially lower than in bulk ($D_M^{(b)} \sim 0.3$ nm²/μs), thus supporting the notion of slower molecular dynamics near the solid surface.

Extension of these studies to properly prepared surfaces and other target fluids could be exploited to directly probe the 'no-slip' condition at the surface, a notion often invoked but notoriously difficult to verify experimentally[28]. Alternatively, polarizing external spins can be a key element in various quantum information protocols using solid state spins, such as in simulating strongly interacting spin systems that can be controlled and readout through single quantum probes[35,36], and for studies on the foundational aspects of quantum thermodynamics.

**Acknowledgements**
The authors acknowledge support from the Max Planck Society, the DFG, ERC Advanced Grant SMeL, BW-Stiftung and BMBF. F.S. expresses the gratitude to Jochen Scheuer for the help with the Qudi software suite installation[33]. C.A.M. acknowledges support from the National Science Foundation through grants NSF-1619896 and NSF-1401632, from the NSF CREST-IDEALS HRD-1547830, and from Research Corporation for Science Advancement through a FRED Award.

**REFERENCES**

[1] K-N. Hu, V.S. Bajaj, M.M. Rosay, R.G. Griffin, "High frequency dynamic nuclear polarization using mixtures of TEMPO and trityl radicals", *J. Chem Phys.* **126**, 044512 (2007).

[2] M.W. Doherty, N.B. Manson, P. Delaney, F. Jelezko, J. Wrachtrup, L.C.L. Hollenberg, "The nitrogen-vacancy colour centre in diamond", *Phys. Rep.* **528**, 1 (2013).

[3] P. London, J. Scheuer, J.-M. Cai, I. Schwarz, A. Retzker, M.B. Plenio, M. Katagiri, T. Teraji, S. Koizumi, J. Isoya, R. Fischer, L.P. McGuinness, B. Naydenov, F. Jelezko, "Detecting and polarizing nuclear spins with double resonance on a single electron spin", *Phys. Rev. Lett.* **111**, 067601 (2013).

[4] R. Fischer, C.O. Bretschneider, P. London, D. Budker, D. Gershoni, L. Frydman "Bulk nuclear polarization enhanced at room temperature by optical pumping", *Phys. Rev. Lett.* **111**, 057601 (2013).

[5] R. Fischer, A. Jarmola, P. Kehayias, D. Budker, "Optical polarization of nuclear ensembles in diamond", *Phys. Rev. B* **87**, 125207 (2013).

[6] D. Pagliero, K.R. Koteswara Rao, P.R. Zangara, S. Dhomkar, H.H. Wong, A. Abril, N. Aslam, A. Parker, J. King, C.E. Avalos, A. Ajoy, J. Wrachtrup, A. Pines, C.A. Meriles, "Multispin-assisted optical pumping of bulk ¹³C nuclear spin polarization in diamond", *Phys. Rev. B* **97**, 024422 (2018).

[7] R. Wunderlich, J. Kohlrautz, B. Abel, J. Haase, J. Meijer, "Optically induced cross relaxation via nitrogen-related defects for bulk diamond ¹³C hyperpolarization", *Phys. Rev. B* **96**, 220407(R) (2017).

[8] G.A. Alvarez, C.O. Bretschneider, R. Fischer, P. London, H. Kanda, S. Onoda, J. Isoya, D. Gershoni, L. Frydman, "Local and bulk ¹³C hyperpolarization in nitrogen-vacancy-centred diamonds at variable fields and orientations", *Nat. Commun.* **6**, 8456 (2015).

[9] J.P. King, K. Jeong, C.C. Vassiliou, C.S. Shin, R.H. Page, C.E. Avalos, H-J. Wang, A. Pines, "Room-temperature in situ nuclear spin hyperpolarization from optically pumped nitrogen vacancy centres in diamond", *Nat. Commun.* **6**, 8965 (2015).

[10] E. Scott, M. Drake, J.A. Reimer, "The phenomenology of optically pumped $^{13}$C NMR in diamond at 7.05 T: Room temperature polarization, orientation dependence, and the effect of defect concentration on polarization dynamics", *J Magn Reson.* **264**, 154 (2016).

[11] B.L. Green, B.G. Breeze, G.J. Rees, J.V. Hanna, J-P. Chou, V. Ivády, A. Gali, M.E. Newton, "All-optical hyperpolarization of electron and nuclear spins in diamond", *Phys. Rev. B* **96**, 054101 (2017).

[12] A.L. Falk, P.V. Klimov, V. Ivády, K. Szász, D.J. Christle, W.F. Koehl, A. Gali, D.D. Awschalom, "Optical polarization of nuclear spins in silicon carbide", *Phys. Rev. Lett.* **114**, 247603 (2015).

[13] A. Ajoy, K. Liu, R. Nazaryan, X. Lv, B. Safvati, G. Wang, D. Arnold, G. Li, A. Lin, P. Raghavan, E. Druga, D. Pagliero, J.A. Reimer, D. Suter, C.A. Meriles, A. Pines, "Orientation-independent room-temperature optical $^{13}$C hyperpolarization in powdered diamond", *Science Adv.*, in press.

[14] J.D.A. Wood, J-P. Tetienne, D.A. Broadway, L.T. Hall, D.A. Simpson, A. Stacey, L.C.L. Hollenberg, "Microwave-free nuclear magnetic resonance at molecular scales", *Nat. Commun.* **8**, 15950 (2017).

[15] A. Henstra, P. Dirksen, J. Schmidt, and W. Wenckebach, "Nuclear spin orientation via electron spin locking", *J. Mag. Reson.* **77**, 389 (1988).

[16] G-Q. Liu, Q.-Q. Jiang, Y-C. Chang, D-Q. Liu, W-X. Li, C-Z. Gu, H.C. Po, W-X. Zhang, N. Zhao, X-Y Pan, "Protection of centre spin coherence by dynamic nuclear spin polarization in diamond", *Nanoscale* **6**, 10134 (2014).

[17] P. Fernandez-Acebal, O. Rosolio, J. Scheuer, C. Mueller, S. Mueller, S. Schmitt, L.P. McGuinness, I. Schwarz, Q. Chen, A. Retzker, B. Naydenov, F. Jelezko, M.B. Plenio, "Towards hyperpolarization of oil molecules via single nitrogen-vacancy centers in diamond", *Nano Lett.*, Just Accepted Manuscript. **DOI:** 10.1021/acs.nanolett.7b05175

[17] T. Staudacher, F. Shi, S. Pezzagna, J. Meijer, J. Du, C.A. Meriles, F. Reinhard, J. Wrachtrup, "Nuclear magnetic resonance spectroscopy on a (5nm)$^3$ volume of liquid and solid samples", *Science* **339**, 561 (2013).

[18] L.M. Pham, S.J. DeVience, F. Casola, I. Lovchinsky, A.O. Sushkov, E. Bersin, J. Lee, E. Urbach, P. Cappellaro, H. Park, A. Yacoby, M.D. Lukin, R.L. Walsworth, "NMR technique for determining the depth of shallow nitrogen-vacancy centers in diamond", *Phys. Rev. B* **93**, 045425 (2016).

[19] *Principles of Magnetic Resonance,* 3$^{rd}$ Ed., C.P. Slichter, Springer Series in Solid-State Sciences, Springer-Verlag, Berlin (2010).

[20] A. Laraoui, C.A. Meriles, "Approach to dark spin cooling in a diamond nanocrystal", *ACS Nano* **7**, 3403 (2013).

[21] A. Henstra, W.Th. Wenckebach, "The theory of nuclear orientation via electron spin locking (NOVEL)", *Mol. Phys.* **106**, 859 (2008).

[22] C.A. Meriles, J. Liang, G. Goldstein, J. Hodges, J. Maze, M.D. Lukin, P. Cappellaro, "Imaging mesoscopic nuclear spin noise with a diamond magnetometer", *J. Chem. Phys.* **133**, 124105 (2010).

[23] N. Aslam, M. Pfender, P. Neumann, R. Reuter, A. Zappe, F. Fávaro de Oliveira, A. Denisenko, H. Sumiya, S. Onoda, J. Isoya, J. Wrachtrup, "Nanoscale nuclear magnetic resonance with chemical resolution", *Science* **357**, 67 (2017).

[24] D. Pagliero, A. Laraoui, C.A. Meriles, "Imaging nuclear spins weakly coupled to a probe paramagnetic center", *Phys. Rev. B* **91**, 205410 (2015).

[25] A. Ajoy, U. Bissbort, M.D. Lukin, R.L. Walsworth, P. Cappellaro, "Atomic-Scale Nuclear Spin Imaging Using Quantum-Assisted Sensors in Diamond", *Phys. Rev. X* **5**, 011001 (2015).

[26] T.M. Staudacher, N. Raatz, S. Pezzagna, J. Meijer, F. Reinhard, C.A. Meriles, J. Wrachtrup, "Probing molecular dynamics at the nanoscale via an individual paramagnetic center", *Nature Commun.* **6**, 8527 (2015).

[27] P.A. Thompson, S.M. Trojan, "A general boundary condition for liquid flow at solid surfaces", *Nature* **389,** 360 (1997).

[28] E. Lauga, M.P. Brenner, H.A. Stone, "Microfluidics: The No-Slip Boundary Condition", Ch. 15 in Handbook of Experimental Fluid Dynamics Editors J. Foss, C. Tropea, A. Yarin, Springer, New-York (2005).

[29] D. Ortiz-Young, H-C. Chiu, S., Kim, K. Voitchovsky, E. Riedo, "The interplay between apparent viscosity and wettability in nanoconfined water", *Nat. Commun.* **4,** 2482 (2013).

[30] D. Abrams, M.E. Trusheim, D. Englund, M.D. Shattuck, C.A. Meriles, "Dynamic nuclear spin polarization of liquids and gases in contact with nanostructured diamond", *Nano Letters* **14**, 2471 (2014).

[31] D.A. Broadway, J-P. Tetienne, A. Stacey, J.D.A. Wood, D.A. Simpson, L.T. Hall, L.C.L. Hollenberg, "Quantum probe hyperpolarisation of molecular nuclear spins", arXiv:1708.05906 (2017).

[32] I.J. Lowe, S. Gade, "Density-matrix derivation of the spin-diffusion equation", *Phys. Rev.* **156**, 817 (1967).

[33] J.M. Binder, A. Stark, N. Tomek, J. Scheuer, F. Frank, K.D. Jahnke, C. Müller, S. Schmitt, M.H. Metsch, T. Unden, T. Gehring, A. Huck, U.L. Andersen, L.J. Rogers & F. Jelezko, "Qudi: a modular python suite for experiment control and data processing", *SoftwareX* **6**, 85-90 (2017).